\documentclass[lettersize,journal]{IEEEtran}
\IEEEoverridecommandlockouts
\usepackage{cite}
\usepackage{amsmath,amssymb,amsfonts}
\usepackage{algorithmic}
\usepackage{graphicx}
\usepackage{textcomp,subcaption}
\usepackage{xcolor,multicol,multirow}
\usepackage{enumerate,textcomp,array,enumitem}
\usepackage{placeins}
\usepackage{threeparttable,url}
\usepackage{dblfloatfix}
\usepackage{flushend}
\usepackage{setspace}
\usepackage{soul}
\usepackage[running,displaymath]{lineno}

\def\BibTeX{{\rm B\kern-.05em{\sc i\kern-.025em b}\kern-.08em
    T\kern-.1667em\lower.7ex\hbox{E}\kern-.125emX}}

\begin{document}

\title{
Synchronous Condensers: Enhancing Stability in Power Systems with Grid-Following Inverters 
}
  
\author{ 
Amir~Sajadi, \IEEEmembership{Senior~Member,~IEEE}, 
Barry~Mather, \IEEEmembership{Senior~Member,~IEEE},
and\\
Bri-Mathias~Hodge, \IEEEmembership{Senior~Member,~IEEE}



\thanks{Authors are with the University of Colorado Boulder and the National Laboratory of the Rockies (NLR), CO, USA.}
}

\markboth{Preprint Draft, \today}%
{Sajadi \MakeLowercase{\textit{et al.}}: TBD}

\maketitle
\begin{abstract} 
Large-scale integration of inverter-based resources into power grids worldwide is challenging their stability and security. This paper takes a closer look at synchronous condensers as a solution to mitigate stability challenges caused by the preponderance of grid-following inverters. It finds that while they are not grid-forming assets themselves, they could enhance grid stability. Throughout this paper, different facets of power system stability and their underlying phenomena are discussed. In addition, instances of instability and mitigation strategies using synchronous condenser are demonstrated using electromagnetic transient simulations. The analysis in this paper highlights the underlying mechanism by which synchronous condensers enhance angular stability, frequency response, and voltage stability. Moreover, it underscores the criticality of their choice of location by demonstrating the destabilizing behavior that could be initiated by the interactions of synchronous condensers.  
\end{abstract}


\section{Introduction}

The ongoing trend of integrating inverter-based resources (IBR) is fundamentally altering the landscape of power system planning and operation \cite{gu2022power}. In practice, almost all utility-scale installations of IBRs which have hitherto been deployed on bulk power systems have employed grid-following inverters (GFLs) \cite{GFM_ESIG_2022}, and only recently grid-forming control technology has begun to be piloted in large-scale systems \cite{Kauai_GFM,SouthAUS_GFM}. 
Despite their widespread installation, it is well documented that instabilities can occur with high penetrations of GFLs mainly due to the supplantation of synchronous generators (SG) \cite{ulbig2014impact,kenyon2023criticality,odessa21,odessa22}. 
This class of inverter control relies on an established voltage and frequency at the point of interconnection to be explicitly tracked with a phase-locked loop (PLL) \cite{lin2020research}. From the circuit analysis perspective, they are a current source with the capability to control active and reactive power by adjusting the current magnitude and power factor, simply tracking the voltage and frequency established at their point of interconnection. Theoretically, operating a power system with 100\% GFLs is infeasible \cite{sajadi2022synchronization} due to their requirement of a waveform with a stale frequency to follow, which explains the impetus behind the stability concern in power grids with high shares of GFL to be practically the low inertia \cite{milano2018foundations} and its implications on voltage and phase angle stability \cite{dozein2025system}. The capability of an asset to create and regulate such a waveform signal is commonly referred to as a grid formation capability. It will be subsequently discussed that the capability to provide frequency and voltage response and the grid-forming capability are two distinct processes, and thus not equivalent.
%

Conventionally, the chief grid-forming assets have been the SGs that also produce active power to serve the load. 
In recent years, as GFLs continue to displace SGs, synchronous condensers (SynCo) have been identified as a promising technology to aid the stability of grids, particularly during operating conditions with high GFL penetration. 
The use of SynCo to enhance power grid operations is a long-standing convention, dating back to the 1920s \cite{alger1928synchronous}. 
As such, the literature addressing SynCo is quite mature. Classically, SGs were retrofited to be used as SynCos post-retirement for voltage regulation and reactive power support purposes, which are more steady-state issues \cite{masood2016post}. However, their utilization for IBR-related stability concerns is relatively new. 

As it relates to the use of SynCos to support IBRs, a number of recent papers have studied the dynamic grid-supporting features of SynCos including voltage support \cite{teleke2008dynamic}, improved frequency RoCoF \cite{nguyen2018combination}, and frequency damping support \cite{nguyen2020applying}. Additionally, it has been shown that SynCos can enhance voltage dynamics and subsequently improve performance in weak grids and preventing sub-synchronous oscillations \cite{hadavi2021robust} or improving Sub/Super-Synchronous oscillation damping \cite{wang2020impact}. It has been also shown that SynCos have the potential to enable the operation of a system that compromises of only GFLs \cite{kenyon2020grid}. 
%
%
%
%
%
%
Similarly, several studies have addressed the application of SynCo in helping resolve system protection issues through an increase in the fault current levels \cite{nedd2017application,liu2023transient} including unbalanced faults \cite{jia2019investigation}, leading to an improved phase angle stability.
%
%
%
%
%
There also have been a number of studies to identify the most suitable locations for SynCos including the cost to benefit ratio mapping for meeting the frequency response obligations \cite{nguyen2021technical} and satisfaction of short-circuit current ratio regional requirements \cite{jia2018synchronous,hadavi2022planning}, including high-voltage direct current (HVDC) lines \cite{wang2024synchronous}.
%
%
%
%
%
Moreover, more recently, parametric evaluation of SynCos has been addressed and the ramifications of design parameters on system-level performance has been studied, including an assessment of SynCo's reactances and excitation systems on the stability of oscillations \cite{bao2024maximizing} and the impacts of SynCo's excitation system and AVR on voltage stability \cite{nair2024parametrically}.

%

%

Based on the the above-mentioned review of the literature, it can safely be concluded that while many capabilities of SynCos in enhancing the stability of power grids with high shares of GFLs have been established, most papers address only individual facets of the multifaceted impact that SynCos can have, leaving a gap in the understanding of the interactive and correlative dynamics of phase angle, voltage, and frequency as a whole. The contribution of this paper is to address this gap by demonstrating the underlying physical phenomena behind the grid stability enhancement that SynCo offer and their interactive dynamics with GFLs. Accordingly, this paper provides an overview of the changing applications of SynCos, departing from their traditional duty as "\textit{power factor correction}" assets and changing into assets for enhancing "\textit{grid strength}" to support the large-scale integration of GFL-IBRs. The findings of this paper establish that while they cannot be be considered a "grid-forming" asset, by their operating principle, they enhance angular stability, frequency response, and voltage stability. Additionally, this paper underscores the criticality of SynCo locations by demonstrating the destabilizing behavior that can be initiated by SynCos oscillating against each other when they are inappropriately sited. Throughout this paper, electromagnetic transient (EMT) simulations of the IEEE 9-bus system in Power Systems Computer Aided Design (PSCAD) are presented to illustrate the points above. This study uses open-source models \cite{kenyon2021open} that available to the public at no cost \cite{kenyon_pypscad_2020}.

\section{Grid-Formation vs. Grid Strength}

\subsection{Scientific Definitions}

We follow a particular distinction in this work, interpreting the term "grid-forming" as \textit{the ability to form the electromagnetic fields between the source and sink}. Therefore, to qualify as a grid-forming asset, it should be able to meet three essential requirements, that are:
\begin{enumerate}
	\item ability to inject power into the grid to act as a source;
	\item ability to independently form and regulate magnetic flux and, subsequently voltage magnitude;
	\item ability to independently form and regulate voltage phase, and subsequently frequency.
\end{enumerate}

We define "grid strength" as the ability to maintain the electromagnetic fields between all sources and sinks across a network and withstand disturbances. This distinction is consistent with the principles of electromagnetism that underlies the transfer of electric energy in the form of electricity and the dynamics that arise from the physical structure of power networks \cite{sajadi2024plane}. 
In a sense, grid formation and grid strength are directly related and the two solutions combined are the complete solution to a classical problem. \textbf{Grid formation is the static solution} where the \textbf{existence of a stable solution} within the feasible region guarantees the electromagnetic link can be formed and power can flow while the system rests at an equilibrium. \textbf{Grid strength is the dynamic solution} where the \textbf{existence of a stable trajectory} within the feasible operational region guarantees the electromagnetic link can withstand the transients, and the flow of power continues as the system transitions between equilibria. This definition is fundamentally consistent with the definition of power stability, defined by the IEEE/CIGRE joint task force \cite{kundur2004definition}: \textit{"Power system stability is essentially a single problem"}. In that definition, they argue that due to the complexity and high dimensionality of the power system stability problem, it needs to be dissected into subproblems in order to be solved and understood. Traditionally, the SGs' principles of operation formed a natural time-hierarchy in power system dynamics which lent itself to an inherent timescale separation in how the frequency, voltage, and angular dynamics manifested themselves. As a result, each of these stability problems were reasonably treated as separate problems. But considering the IBRs' principles of operation, such timescale separation no longer exist in the same manner. Therefore, it is becoming increasingly necessary to treat the power system stability problem as one, as defined in \cite{kundur2004definition}, which in essence is the "\textit{grid strength}" problem.

\subsection{Asset To "Form" the Grid?}
\label{sec:forming}

\subsubsection{GFL-IBR}

GFLs are current sources that rely on matching a voltage waveform at their point of interconnection, thus they are not grid-forming assets. This can be interpreted as current injection when voltage is provided by the grid. The basics of electrical engineering suggests that while both voltage and current are essential in the transfer of energy (and inevitably power transfer), voltage is more fundamental than current in circuit analysis. Voltage is formed by the presence of electric charges that create potential differences, whereas current is formed by the flow of electric charges. From this perspective, we argue that GFLs then contribute to AC power transfer by creating continuous disturbances in their host electromagnetic field as the impetus behind the vibration of electric charges, i.e. current injection in circuit analysis. Nevertheless, the prerequisite is the availability of a certain amount of electric charge that will form the voltage level before they can be disturbed, that is the grid formation capability. The GFL lacks such capability.

\subsubsection{Synchronous Condenser}

A SynCo is a motor that operates with the field current being overexcited, injecting reactive current into the grid; thus it acts as a variable capacitor. Because of this injection, it meets the first requirement. The next question to discuss is whether the injection of reactive power simultaneous to the draw of active current constitutes grid formation or not. 

For the voltage magnitude, it is true that the doubly excited nature of SynCos offers an exquisite independent capability to form voltage at its terminal even at no load conditions; that is the excitation (internal) electromotive force (emf). In SGs the prime mover to form mechanical power is provided by the turbine, whereas in SynCo it is provided by the host network in the form of active power. Regardless of the form of prime mover, the voltage magnitude can be formed independently and controlled by field current. 

As for the voltage phase angle and frequency: if a SynCo is directly connected to its host network, at nominal speed operation it spontaneously synchronizes with the frequency of the network from which its draws the power that allows it to rotate. This indicates the lack of an independent ability to form voltage phase angle and frequency, and instead it follows the nodal condition of the hosting grid. We argue that it could be interpreted as a mechanical analogue of the function that a PLL performs in a GFL. An alternative to this setting could be to run the synchronous condenser with an industrial high-power drive that provides speed control, e.g. a variable frequency drive. This would remove the direct electromagnetic coupling of the synchronous condenser and the network, making the synchronous condenser an asynchronous apparatus. As a result, it may eliminate the benefits that a synchronous condenser may provide to enhancing the dynamics of its host grid, a subject that we will discuss in the next section. Studying the operation of an asynchronous doubly excited condenser is beyond the scope of this work.

\subsubsection{Summary}

The most important conclusions of our discussion above are that neither GFL nor SynCos meet the constituting criteria for a grid-forming asset. In a sense, a SynCo can be viewed as a mechanical GFL with the ability to form voltage magnitude and the operational benefits that comes with its mechanical construct, as we will discuss in the next section.

\subsection{Asset to Provide "Strength" To the Grid?}


\subsubsection{GFL-IBR}


As discussed in the preceding section, GFLs are current sources and unable to independently form voltage magnitude or phase. Therefore, frequency response and voltage support provided by GFLs are consequences of regulation of active and reactive power to support the grid during transients. The phase angle, $\delta_{\text{PLL}}$, and subsequently, frequency, $\omega_{\text{PLL}}$ are estimated by the PLL, and the terminal voltage, $v_g$ is directly measured. Then the active and reactive power set points, $P_{set}$ and $Q_{set}$, are regulated as $P^*$ and $Q^*$, respectively, according to the changing grid conditions when deviations from the nominal frequency and voltage, $\omega_n$ and $v_n$, are registered by the PLL and terminal measurement and consistent with the respective droop for active and reactive power control, $m_p$ and $m_q$, as follows:
\begin{align} 
\begin{split} \label{eq:GFL_PQ}
P^*&=P_{set}+(\omega_n-\omega_{\text{PLL}}) m_p \\
Q^*&=Q_{set}+(v_n-v_g) m_q
\end{split}
\end{align} 

The active power is the rate of work in the field that contributes to the energy propagation, whereas the reactive power could be interpreted as the destructive interference in the total energy flux that is caused by a phase shift between voltage and current, producing the power that oscillates back and forth along the network. 
In the electromagnetic interpretation, the dynamic support from a GFL with grid-supporting capabilities is twofold. 
\begin{itemize}
	\item First, it reduces the stress on the electromagnetic link through the injection of active power that is understood as a form of frequency response. In effect, it acts as a controllable negative load with dynamic response. 
	\item Second, it reduces the destructive interference by supplying additional reactive power in the form of voltage response that, as a secondary effect, may contribute to the adjustment of the shift between voltage and current, thus helping increase the efficiency of energy transfer in the network. 
\end{itemize}

In essence, the grid-supporting functionalities of GFLs are simply alleviating the stress on the network by regulation, but not formation. Additionally, during fault operation, GFLs have a very limited contribution towards short-circuit current because of the limitations of their semi-conductor switches.

\subsubsection{Synchronous Condenser}


In a SynCo the electric phase angle, $\delta_{\text{SynCo}}$, is determined by the velocity at which the magnetic field in the SynCo air gap rotates, $\omega_{\text{SynCo}}$, that is directly proportional to the mechanical velocity of the rotor. Its dynamics are described by the swing equation as \cite{krause2002analysis}:
\begin{align} 
\begin{split} \label{eq:swing_basic_delta}
\frac{d^2 \delta_{\text{SynCo}}}{dt^2}  = \frac{d \omega_{\text{SynCo}}}{dt} = M^{-1}   \Big( P_e - P_l - D   \frac{d \delta_{\text{SynCo}}}{dt} \Big)
\end{split}
\end{align} 
where $M$ is the SynCo's angular momentum and $D$ is its damping coefficient. $P_e$ and $P_l$ are electric power input and mechanical power consumption of the load, respectively. 
For a SynCo, given that it operates free-running, $P_l=0$, therefore, the dynamics of phase angle and frequency can be described by: 
\begin{align} 
\begin{split} \label{eq:swing_SC_delta}
\frac{d^2 \delta_{\text{SynCo}}}{dt^2}  = \frac{d \omega_{\text{SynCo}}}{dt} = M^{-1} \Big( P_e  - D  \frac{d \delta_{\text{SynCo}}}{dt} \Big)
\end{split}
\end{align} 

Eq. \eqref{eq:swing_SC_delta} explains that when a SynCo is subjected to an external disturbance where $P_e$ is changed - for example when there is a fault on the network, the $P_e$ drops - the SynCo does not immediately come to a full stop. Instead, it continues to rotate for a short period of time, up to a few seconds, mainly because of the kinetic energy stored in the rotor's spinning shaft; it acts as a generator before it returns to motor mode once it reaches a steady state. The damping coefficient in a synchronous motor is generally a very small value and negligible; hence we make an assumption henceforth $D\approx 0$, simplifying the phase angle and frequency dynamics to: 
\begin{align} 
\begin{split} \label{eq:swing_SCc_delta_simplified}
\frac{d^2 \Delta \delta_{\text{SynCo}}}{dt^2}  = \frac{d \Delta \omega_{\text{SynCo}}}{dt} = M^{-1} \cdot \Delta P_e 
\end{split}
\end{align}  
where $\Delta$ is the linear operator. Integrating \eqref{eq:swing_SCc_delta_simplified} with respect to time gives

\begin{align} 
\begin{split} \label{eq:swing_SCc_delta_simplified_int1}
\Delta f_{\text{SynCo}} (t) = \frac{d \Delta \delta_{\text{SynCo}}}{dt}  = (2\cdot \pi \cdot M)^{-1} \cdot \Delta P_e \cdot t
\end{split}
\end{align}  
A second integration produces an electric angle trajectory as a function of time:
\begin{align} 
\begin{split} \label{eq:swing_SCc_delta_simplified_int2}
\Delta \delta_{\text{SynCo}} (t)   = M^{-1} \cdot \Delta P_e \cdot t^2 + \delta_{{\text{SynCo}}_0}
\end{split}
\end{align} 
For the duration of the external disturbance, a SynCo slows down with time and eventually may stop, since it has become a source to the grid without a prime mover to support its continuous operation. 
From Eqs. \eqref{eq:swing_SCc_delta_simplified_int1} and \eqref{eq:swing_SCc_delta_simplified_int2} we understand that this kinetic energy is directly proportional to the $M$ momentum value, that is interchangeably and colloquially referred to as inertia in industry; the larger the momentum (or inertia), the longer the rotor will continue to spin as it slows down, therefore, the longer the phase angle continues to evolve and the slower the frequency decline. Eqs. \eqref{eq:swing_SCc_delta_simplified_int1} and \eqref{eq:swing_SCc_delta_simplified_int2} also show that in case of an electric fault the longer the clearance time, reflected by $t$ in these equations, the larger the frequency and phase angle deviation. Therefore, the shorter the clearance time, the better. 
It should be noted that a SynCo is not a self-starter machine, thus in case of significant slow down or a full stop, it will need auxiliary support to restart. This is a major drawback for SynCos, not having inherent blackstart capability.

Immediately following the inception of a fault short-circuit currents inside a SynCo are induced. They can be understood using the \textit{constant flux linkage theorem}. Similar to the description provided in \cite{machowski2008power} of short-circuit currents for SG, the flux linkage inside a SynCo is produced jointly by rotating flux that excitation current drives and, $\Psi_{f}$, the flux that the armature current drives, $\Psi_{a}$. As a disturbance occurs, according to the theorem of constant flux linkage, the flux linkages inside the SynCo cannot immediately change and instead should hold their exact value at the instance of fault occurrence, that would be $\Psi_{f_{0^{-}}}$ and $\Psi_{a_{0^{-}}}$. After the disturbance, assuming it is a short-circuit fault, the voltage drop at a SynCo terminal causes the electric power delivered from the network to the machine to drop. However, the SynCo shaft continues to rotate, because of its inertia provided by the kinetic energy it stores, effectively turning itself briefly into a SG. As a SynCo continues to rotate, the driving fluxes also change because of the change in driving current, $\Psi_{f_{0^{+}}}$ and $\Psi_{a_{0^{+}}}$. Applying the theorem of constant flux linkage, in order to maintain a constant flux linkage in each phase, additional currents need is induced in the windings to make up for the necessary flux. We denote the flux linkage driven by the field transiently induced current $\Psi_{f_\gamma}$ and the part driven by the armature transiently induced current $\Psi_{a_\gamma}$. The overall transient flux linkage for each phase winding can be computed by combining all components \cite{machowski2008power}:
\begin{align} 
\begin{split} \label{eq:dq_power_linear}
\Psi_{f_{0^{-}}} & = \Psi_{f_\gamma} + \Psi_{f_{0^{+}}} \\
\Psi_{a_{0^{-}}} & = \Psi_{a_\gamma} + \Psi_{a_{0^{+}}} 
\end{split}
\end{align} 
Since $i=\frac{\Psi}{L}$, then  the induced current is a function of SynCo inductance; the larger the SynCo, the larger the transiently induced currents. The stored magnetic energy decays with time due to winding resistance and induced transient currents disappear. 

During a fault, the additional currents force the flux to take a different magnetic path than in the steady state, because of the screening effect, called the transient state \cite{machowski2008power}. Therefore, during this time it exhibits different magnetic characteristics and, subsequently, electrical characteristics. This gives rise to subtransient and transient reactances of the SynCo, $X''_d$ and $X'_d$ along $d-$ axis and $X''_q$ and $X'_q$ along the $q-$ axis. which are often much smaller than steady-state reactances, that are $X_d$ and $X_q$ along the $d-$ and $q-$ axis, respectively. It should be noted that even though the voltage at a SynCo's terminal drops below the nominal value, the short-circuit current will flow towards the fault location which naturally will have a lower voltage. 

Given the characteristics of SynCo explained above, one can see how they are capable of supplying transient short-circuit current that contains active and reactive components. Additional currents induced within the SynCo and smaller reactances that are in effect during a fault cumulatively contribute to the large short-circuit current supplied by SynCos during a fault for a short period of time.


The longstanding feature of SynCos that has appealed to system operators is their ability to regulate power factor through a wide range of reactive power consumption/generation. This makes it evident that with an active excitation system, the SynCo should be able to relatively easily regulate steady-state voltage in transmission networks. Furthermore, earlier in this section, we discussed the capabilities of a SynCo during transient operation with a focus on short-circuit current. Now, let us examine the capabilities of a SynCo to provide voltage transient response and damp out voltage oscillation. To this end, we shift our attention to the dynamic model of armature reaction and the excitation system. 

The steady state model of a SynCo \cite{machowski2008power} assumes the armature reaction emf along the $d-$ axis, denoted by $e_d=0$, and the armature reaction emf along $q-$ axis and the excitation emf, that are denoted by $e_f$ and $e_q$, respectively, have the same interpretation. But during a disturbance this no longer holds. During a disturbance, $e_d$ may no longer be zero and $e_q$ may vary from $e_f$ as a function of changes in field current. Similarly, $e_f$ may change as a function of excitation control. IEEE Std. 666 \cite{IEEE_666_2007} defines the SynCo as current sources therefore they should be modeled in its subtransient state. Accordingly, the dynamics of a SynCo's armature reaction emf during subtransient state described in \eqref{eq:armature_react} \cite{machowski2008power,sauer1998power}, assuming a sufficiently short period of time following the large disturbance that SynCo has transitioned to generator mode, before it returns to motor mode once it reaches an equilibrium.
\begin{align}
\begin{split} \label{eq:armature_react}
\dot{e}''_q & = \big(\dfrac{1}{T''_{do}}\big) \Big(- e''_q + (X'_d - X''_d) \cdot  i_d + e'_{q}\Big)                  \\
\dot{e}''_d &= \big(\dfrac{1}{T''_{qo}}\big)    \Big( - e''_d - (X'_q - X''_q) \cdot i_q   - e'_{d}\Big)             
\end{split}
\end{align}
where $e''_d$ and $e''_q$ are subtransient emfs and $e'_d$ and $e'_q$ are transient emfs, all along $d-$ and $q-$ axis. $T''_{do}$ and $T''_{qo}$ are time constants of subtransient state, corresponding to the duration that it takes for the transiently induced currents to decay. $i_d$ and $i_q$ are stator current components along $d-$ and $q-$ axis, respectively. The time constants are a function of $\frac{X}{R}$, indicating the higher the $\frac{X}{R}$ of a SynCo the more robust its voltage transients. During subtransient operation, the $X$ will be the subtransient reactance. In a sense, relationship between the transient voltage time constant, $\frac{X}{R}$, to voltage oscillations can be viewed as the analogue of inertia time constant, $H$ to frequency oscillations.

\section{Simulation Methodology}


\subsection{Test Case}

Electromagnetic transient (EMT) simulation of the Western System Coordinating Council (WSCC) 3-generator, 9-bus system (shown in Fig. \ref{fig:PSCAD_sim_case}) as a test case (also known as the Anderson 9-bus \cite{sauer1998power}) was used to demonstrate the concepts discussed. Our models were built on the foundation of a set of open-source PSCAD models previously developed by NREL \cite{kenyon2021open,kenyon_pypscad_2020}. All generators were rated at $200 \text{MVA}$. All loads were modeled as constant power with no frequency or voltage dependence. Synchronous generators were modeled using generic components available in the PSCAD standard library with an inertia constant of $\text{H}=4\text{s}$. GFL-IBRs were modeled using their full-order representation that includes the inner current loops and associated filter elements. SynCos were also modeled using generic components available in the PSCAD standard library. The exciter was of the solid-state type (ST1A) with voltage measurement at the generator bus to which it is connected and no governor was installed. 
 
\subsection{Modeling Implementation}

The first goal of the simulations was to evaluate the SynCo's capability to operate as a "grid-forming" asset. To this end, we built a case with all three generators, at nodes 1 through 3, being GFL-IBR. Each GFL-IBR was paired with a SynCo. To initiate the simulation, we installed an auxiliary startup synchronous generator as an ideal source to start the system. Then the auxiliary generator's breaker was opened once the system reaches a stable steady-state to assess whether the system with all GFL-IBRs and SynCos will be self-sustaining.

\begin{figure}[h]   
	\centering  
	\vspace{-1em}
	\includegraphics[width=.6\linewidth]{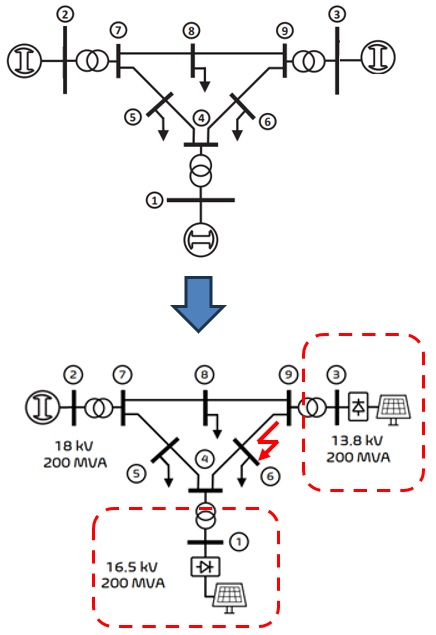}   
	\caption{One-line diagram of the case study}
	\label{fig:PSCAD_sim_case} 
\end{figure}

As a second goal of these simulation, to assess the grid strength enhancement capabilities of SynCos, we defined the base case scenario as a case with two GFL-IBRs and one synchronous generator. We made this decision based upon an observation from our recent study that produced the tutorial on GFL-IBR integration \cite{sajadi2023dynamics}. In the former study, we incrementally supplanted synchronous generators with GFL-IBRs and determined that cases where two out of the three generators were GFL-IBRs are extremely susceptible to instabilities. Accordingly, in the present study, we arbitrarily selected a case with GFL-IBRs supplying power as generators at buses 1 and 3 while bus 2 hosts a synchronous generator, (depicted in Fig. \ref{fig:PSCAD_sim_case}), 
all with equal capacity of 200 MVA (that corresponds to case GFL13 in \cite{sajadi2023dynamics}). We subjected this system to a three-phase balanced electrical short-circuit event at bus 6 (highlighted in Fig. \ref{fig:PSCAD_sim_case}) 
that was cleared after $0.083\text{s}$, equal to $5$ electrical cycles. Measurements from the generator attached to bus 3 are shown throughout the analysis. To ensure the generality of the inferences we make, we carefully reviewed signals from the other generators and they were conclusively in agreement with the observations shown throughout the rest of this paper.

\section{Results and Discussion}

\subsection{Grid-forming Capability}

\begin{figure}[h]   
	\centering  
	\includegraphics[width=.6\linewidth]{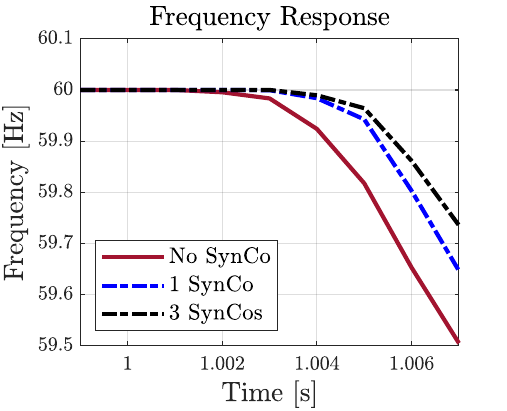}   
	\caption{Frequency stability for three cases examined for with all GFL-IBRs}
	\label{fig:fig_freqs_All_GFL} 
\end{figure}

To evaluate the SynCo's capability to operate as a "grid-forming" asset, we ran the case with all GFL-IBRs accompanied by an auxiliary startup generator. Once the system reached a steady-state, the breaker for the auxiliary generator was opened. This resulted in a prompt system frequency instability, highlighted by the red solid trace in Fig. \ref{fig:fig_freqs_All_GFL}. This is mainly because once the breaker opens, the PLL did not have a stable reference frequency signal to track due to the lack of a grid-forming asset in the network; thus the GFL-IBRs shut down.

We repeated this test by adding a SynCo with a capacity of $S_1=14.85  \text{MVA}$ at node 1. This experiment also resulted in frequency becoming unstable nearly instantly after the breaker opened, but with a small delay relative to the experiment with no SynCo, as distinguished by the blue dashed trace in Fig. \ref{fig:fig_freqs_All_GFL}. Finally, we installed three SynCos, one paired with each GFL-IBR, with capacities of $S_1=14.85 \text{MVA}$, $S_2=14.58  \text{MVA}$, and $S_3=20.70  \text{MVA}$. The results, plotted in the dashed black line in Fig. \ref{fig:fig_freqs_All_GFL}, indicated a swift frequency collapse, with a scarcely detectable difference relative to the previous case. The frequency instabilities in cases with SynCos were caused by the lack of a grid forming asset in the network once the auxiliary generator trips; PLLs have no stable reference frequency signal to track and the SynCos follow the unstable frequency of the network. As a result, the GFL-IBRs ceased operation and the SynCos shut down. 

In the initial case with only GFL-IBRs (labeled as "No SynCo"), the frequency dropped to the under-frequency load shedding (UFLS) threshold, $59.50\text{Hz}$, within $0.007\text{s}$ following the disconnection of the auxiliary generator. That is approximately $0.42$ electrical cycles, whereas in the two subsequent cases with one and three SynCos, within the same time period, frequency drops to $59.65\text{Hz}$ and $59.73\text{Hz}$, respectively. The slow down in frequency decline is because of the mechanical inertia provided by the SynCo(s); the higher the inertia, the longer the delay before instability transpires. Our results clearly show the lack of grid-forming capabilities for paired GFL-IBRs and SynCo systems.

\subsection{Grid Strength Enhancement}

\begin{figure}[h]   
	
	\centering  
	\includegraphics[width=.75\linewidth]{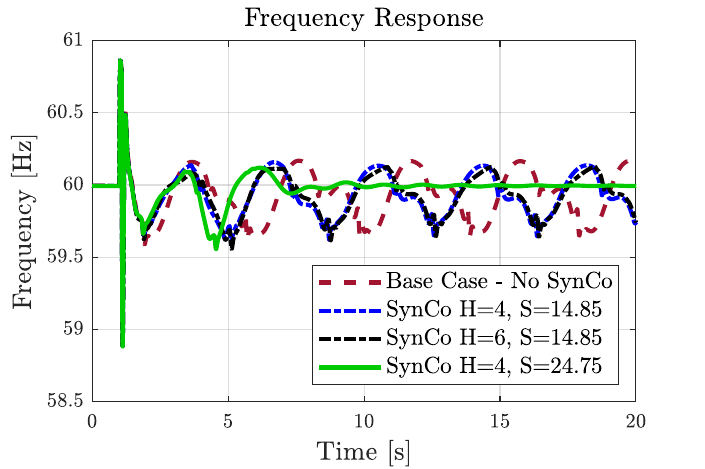}   
	\caption{Frequency response for four cases examined for the installation of SynCo - the pattern of motions in voltage and phase angle signals for each case were similar to those of frequency, thus not shown here.}
	\label{fig:freqs_C1}   
\end{figure}

For the grid strength enhancement analysis we subjected the system to a short-circuit fault on node $6$ with clearance after $5$ electrical cycles. The results for the base case (labelled as "No SynCo" in the plots) was as expected; instability was observed in the form of periodic motions that produces sustained oscillations of all variables, e.g., phase angles, frequency, and voltage. 
To mitigate the instability we added a single SynCo paired with a GFL-IBR at bus 1, rated at $14.85 \text{ MVA}$, and with inertia constant of $\text{H}=4\text{s}$, which appeared insufficient as the instability remained. 
Our attempt to stabilize the system by increasing the inertia constant to $\text{H}=6\text{s}$ while keeping the machine rating at $\text{S}=14.85 \text{ MVA}$ was also unsuccessful.  
However, increasing the machine rating to $\text{S}=24.75 \text{ MVA}$ with the inertia constant being kept at the initial $\text{H}=4\text{s}$ stabilized the system.  
Results are shown in Fig. \ref{fig:freqs_C1}.

These results indicate that, with the addition of a SynCo, the system experiences improved dynamic performance, for all three cases examined. In the two cases where the condenser rating was $\text{S}=14.85\text{ MVA}$ (distinguished by the blue and black traces, respectively, in Fig. \ref{fig:freqs_C1}) the system was not stabilized, but the period of sustained oscillations was shortened relative to that of the base case (depicted by the red traces in Fig. \ref{fig:freqs_C1}), suggesting an improved damping-like effect. Increasing the SynCo rating to $\text{S}=24.75\text {MVA}$ (distinguished by the green traces in Fig. \ref{fig:freqs_C1}) substantially improved the dynamic performance, resulting in a stable response.

\begin{figure}[h]    
	\centering  
	\includegraphics[width=.75\linewidth]{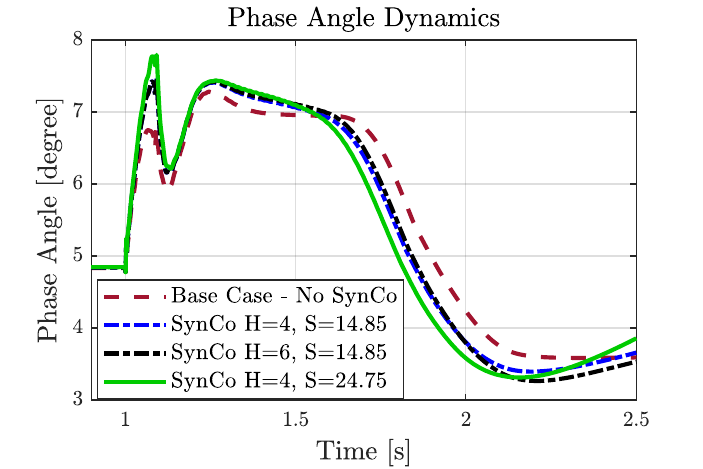}   
	\caption{Phase angle fast dynamics for four cases examined for the installation of SynCo}
	\label{fig:fig_angs_gen_r_C1_transients}    
\end{figure}

The fact that additional inertia did not stabilize the system but additional capacity did presents circumstantial evidence that this instability could have been caused by either phase angle or voltage issues. To better understand the root cause of this instability, we restrict our attention to system fast transients during the fault and within the first few cycles following its clearance. After a careful review of the frequency, phase angle, and voltage fast transients 
the results indicate that the root cause of this instability was phase angle instability, because the installation of a SynCo with a larger nominal rating enhanced the phase angle stability. Higher inertia did not make a considerable impact. It is evidenced in Fig. \ref{fig:fig_angs_gen_r_C1_transients} as a larger angle peak during the fault for the case with higher capacity and an no change in phase angle response for the case with higher inertia. Frequency response and voltage fast transients (during the fault) appear to follow very similar trajectories with negligible differences, thus are not shown. We argue that the use of a larger SynCo alleviated the stress from the only synchronous generator in the system, allowing it to provide a better regulation of phase angle and, thus, more stable frequency.

\begin{figure}[h]   
	\centering
	\begin{subfigure}[b]{0.24\textwidth}
		\centering		\includegraphics[width=\linewidth]{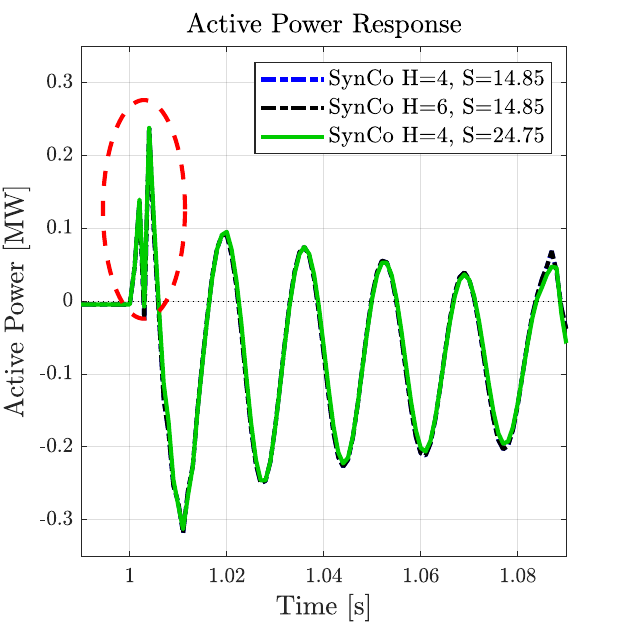}  
		\caption{Active power response}
	\end{subfigure}
	\begin{subfigure}[b]{0.24\textwidth}
		\centering		\includegraphics[width=\linewidth]{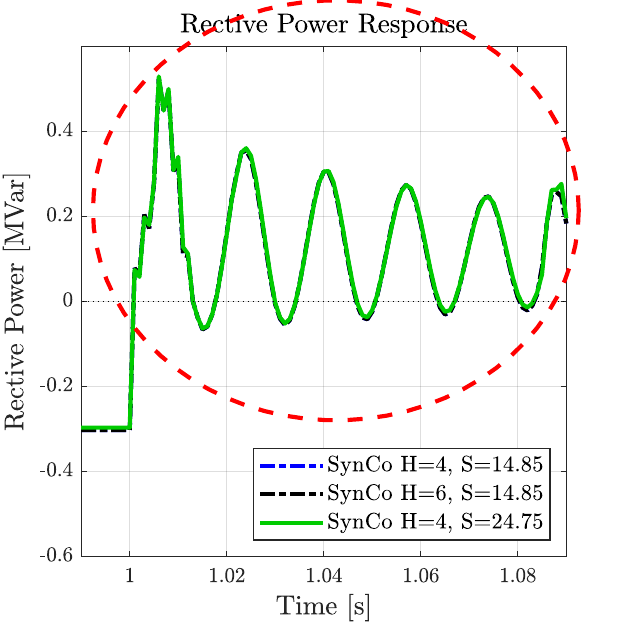}  
		\caption{Reactive power response}
	\end{subfigure}    
	\caption{Active and reactive power response of SynCos - the section highlighted by dashed red circle is the short-circuit power supply}
	\label{fig:fig_PQs_syncos}  
\end{figure}

An important feature of SynCos is their ability to provide short-circuit current in the form of both active and reactive power. The power profiles of the SynCos in the three cases studied are shown in Fig. \ref{fig:fig_PQs_syncos}.

\begin{figure}[h]    
	\centering  
	\includegraphics[width=.75\linewidth]{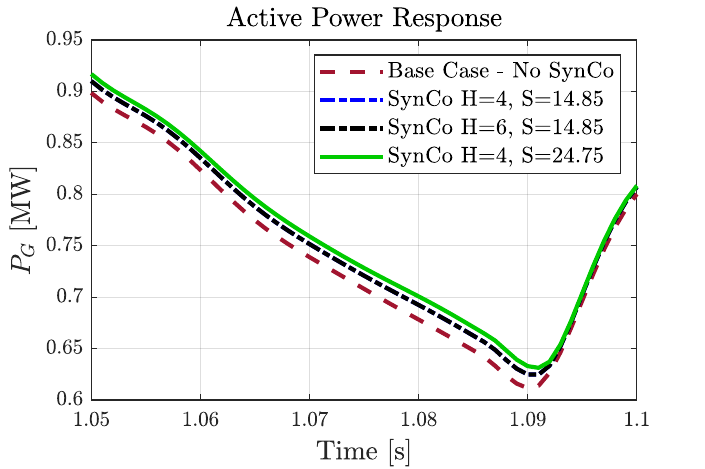}   
	\caption{Active power transient response of synchronous generator at node 2 during the fault}
	\label{fig:fig_Pgs_C1_transients}  
\end{figure}

It can be seen that prior to a fault the system is in a visible steady state, therefore active and reactive flows are at rest. Once a fault occurs, due to the change in network structure, the flow of power instantaneously changes. This is reflected in the change of power supply by the generators and SynCos. The magnitude of change in power flows is a function of the severity of the fault. The characteristics of oscillations that may be present following the clearance of the fault are determined by the structural properties of the system and its ability to remain within the stable region during the fault. As discussed earlier in Section \ref{subsection:SC-SCC}, a SynCo temporarily becomes a synchronous generator, usually for the entirety or a fraction of the fault period. Therefore, it supplies short-circuit current in the form of injection of active and reactive power during the fault, as shown in \ref{fig:fig_PQs_syncos}. In these plots, the injection of power is recognized by the positive quantities of power. Standing out very visibly is the short-circuit power injection by the SynCos in all three cases. A closer look indicated the higher the capacity of the condenser, the higher the amount of active and reactive power injection during the fault. Therefore, the case that has a SynCo with the highest rating, $\text{S}=24.75 \text{ MVA}$, was stable and the two other cases with a lower rating of $\text{S}=14.85 \text{ MVA}$ were not. The inertia appeared to be inconsequential in determining the amount of short-circuit power. The additional active and reactive power support from the SynCo was just enough to relieve the only synchronous generator in the system, allowing it to better regulate phase angle stability, as evidenced by the results shown in Fig. \ref{fig:fig_Pgs_C1_transients}.

In this plot, it can be seen that in the case with the large SynCo (highlighted by the solid green line), the synchronous generator injects slightly less active power during the fault. Subsequently, its rotor angle reaches the boundary of stability but does not exit it. Hence, the system remains stable.

A widely used metric to assess the impact of SynCos is the improvement of the $\frac{X}{R}$ ratio at its point of interconnection. This metric quantifies the short-circuit current, according to IEEE Std. 666 \cite{IEEE_666_2007} and IEEE Std. C37.010 \cite{IEEE_C37_010_2016}. In the base case with two GFL-IBRs, that are connected to nodes 1 and 3, given their extremely limited capabilities to provide short-circuit current, the $\frac{X}{R}$ at the GFL-IBR nodes is nearly zero and the entire short-circuit current in the system is supplied by the only synchronous generator in the system, that is connected to node 2. Hence, the $\frac{X}{R}$ is only that of that generator.
In the two subsequent cases with a SynCo, the condenser is the main source of $\frac{X}{R}$ at its interconnection node where it is paired with a GFL-IBR. In other words, the installation of a SynCo naturally improves the $\frac{X}{R}$ ratio because of their supply of short-circuit power. This metric is particularly prevalent in areas where voltage oscillations are an issue. 

\begin{figure}[h]    
	\centering  
	\includegraphics[width=.75\linewidth]{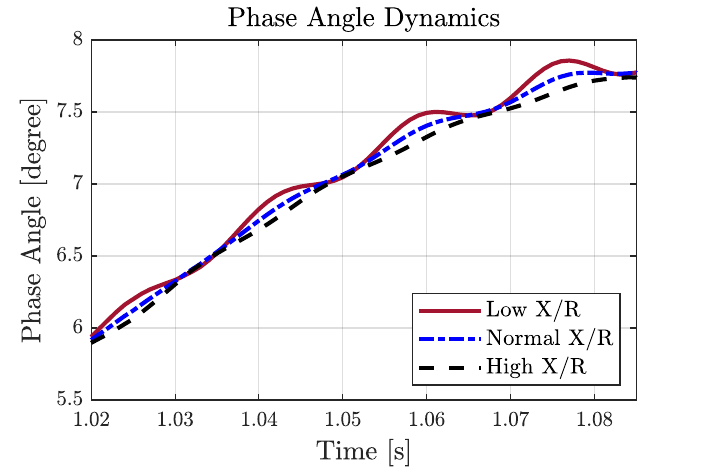}   
	\caption{Phase angle fast transients for different $\frac{X}{R}$ ratios}
	\label{fig:fig_angs_gen_r_XR_transients}  
\end{figure}

According to IEEE Std. 666 \cite{IEEE_666_2007}, SynCos are considered as current sources, and therefore are modeled using their subtransient reactance, $X^{''}_d$. However, this reactance is an inherent property of the machine design and cannot be algorithmically adjusted. Nonetheless, for the sake of an exploratory assessment, in the latter case with a SynCo connected to node 1, rated at $S_1=24.75 \text{ MVA}$ and inertia time constant of $\text{H}=4\text{s}$, we slightly varied the $\frac{X}{R}$ by first decreasing its subtransient reactance from $0.220 \text{p.u.}$ to $0.150 \text{p.u.}$ and, subsequently, increasing it to $0.295 \text{p.u.}$. After a careful review of the results, the main impact was identified in phase angle oscillations, as shown in Fig. \ref{fig:fig_angs_gen_r_XR_transients}. These oscillations were also mildly visible in frequency response.

These results indicate that a lower level of $\frac{X}{R}$ could introduce oscillations in the phase angle, as highlighted by the red solid trace. In an area with poor voltage damping characteristics, that is generally referred to as a "weak grid", these oscillations may appear in the voltage signal too. Overall, we postulate that the use of $\frac{X}{R}$ as a metric in an area prone to oscillations simply is an indicative of the need for the installation of a synchronous machine, either generator or condenser, that in addition to providing voltage damping (which is inherent to the function of an exciter), may fix other underlying stability issues, e.g., phase angle and frequency response. The traces of frequency and phase angle for the three cases follow similar trajectories, thus are not shown here.

\begin{figure}[h]    
	\centering  
	\includegraphics[width=.75\linewidth]{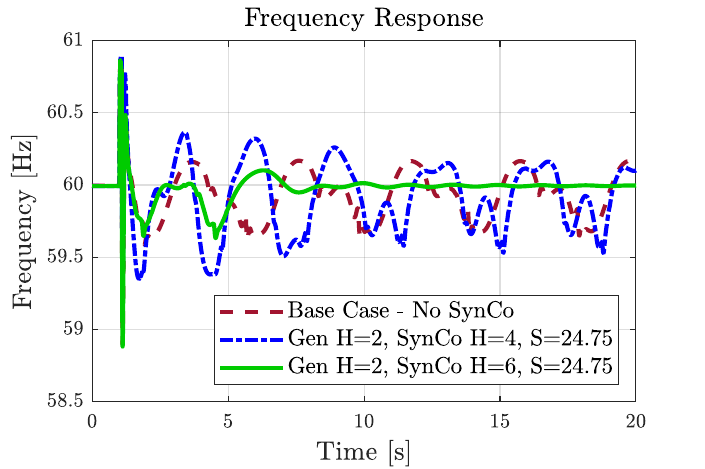}   
	\caption{Frequency response for cases examined for the system with reduced inertia - the pattern of motions in voltage and phase angle signals for each case were similar to those of frequency, thus not shown here.}
	\label{fig:freqs_C1_genH}   
\end{figure}

To demonstrate the criticality of the support provided by the inertia of a SynCo, we modified the last case which was stable (labelled as SynCo, $\text{H}=4\text{s}$, $\text{S}=24.75\text{MVA}$) and reduced the inertial constant of the only synchronous generator in the system - that is interconnected to bus 2 - from $\text{H}=4\text{s}$ to $\text{H}=2\text{s}$ without any parametric changes to the SynCo. This change in the generator's inertia destabilized the system again. As a remedy, an increase of the SynCo's inertia from  $\text{H}=4\text{s}$ to $\text{H}=6\text{s}$ while keeping the rating unchanged proved successful in stabilizing the system again. The results are shown in Fig. \ref{fig:freqs_C1_genH}.

These results show that, compared to the base case (depicted by the red traces in Fig. \ref{fig:freqs_C1_genH}, the addition of the SynCo improved the system's dynamic performance even when it did not fully stabilize the system. The improvement manifests itself in the form of shorter periods of oscillations, regardless of whether these motions are sustained, as was in the former case with the SynCo's inertia constant $\text{H}=4\text{s}$ (shown by the blue traces in Fig. \ref{fig:freqs_C1_genH}, or damped, as was in the latter case with the SynCo's inertia constant $\text{H}=6\text{s}$ (distinguished by the green traces in Fig. \ref{fig:freqs_C1_genH}).

\begin{figure}[h]   
	\centering  
	\includegraphics[width=.75\linewidth]{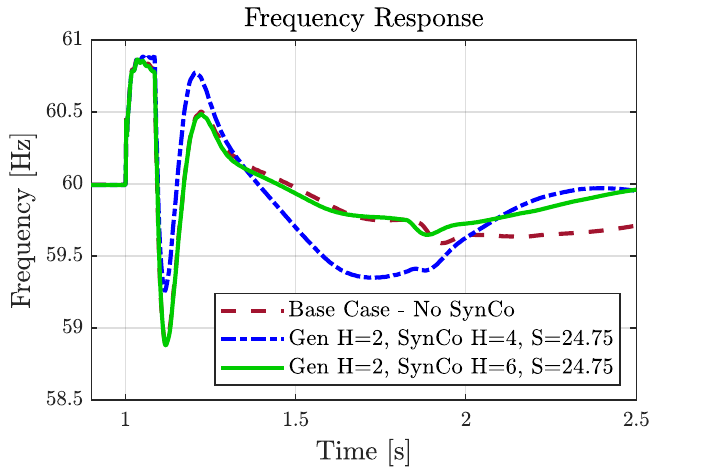}   
	\caption{Frequency fast dynamics for cases examined with reduced inertia}
	\label{fig:fig_freqs_C1_transients_genH}   
\end{figure}

To better understand the contribution provided by a SynCo in low-inertia scenarios, we narrow our attention to the system's fast dynamics. 
After a careful review of the results for phase angle, voltage, and frequency fast transients, frequency emerged as the primary contributing factor, shown in Fig. \ref{fig:fig_freqs_C1_transients_genH}.

These results clearly indicate that the installation of a SynCo with higher inertia helped produce a better frequency response, highlighted by the green trace, which directly corresponds with the root cause of this instability, which was low inertia. Phase angle and voltage fast transients did not appear as determining factors in this case, and hence are not shown.

Our analysis concludes that the selection of a SynCo and its optimal characteristics, e.g. capacity, inertia, and $\frac{X}{R}$ ratio, needs to be specific to the network phase angle, frequency, and/or voltage challenges and needs. We conducted these experiments for more cases where the only SynCo in the system was paired the other GFL-IBR at bus 3, and the results were similar, thus are not shown.

\subsection{Multi-Synchronous Condenser}

\begin{figure}[h]   
	\centering   
	\includegraphics[width=.75\linewidth]{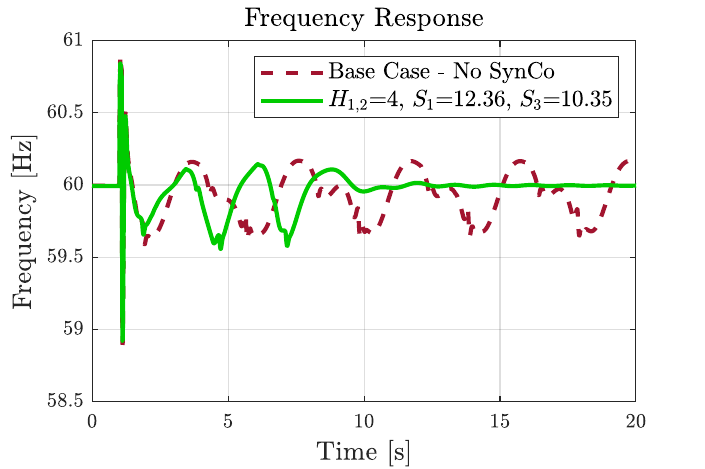}   
	\caption{Frequency response for the case with two SynCo - stabilizing the base case}
	\label{fig:double_base} 
\end{figure}

Now we turn our attention to systems with more than one SynCo and how they may interact. To this end, we installed two SynCos in the base case with an inertia constant of $\text{H}=4{s}$ to stabilize the base case; a SynCo, rated at $\text{S}=12.36\text{ MVA}$, paired with GFL-IBR at bus 1 and a second SynCo, rated at $\text{S}=10.35 \text{ MVA}$, paired with the GFL-IBR at bus 3. 
The results are shown in Fig. \ref{fig:double_base}.
Note that given the similar pattern of oscillations in frequency, phase angle, and voltage that we observed thus far, henceforth, we only show frequency response in the discussion of the dynamic performance of a system.

These results show that in a case with two SynCos the frequency came to a steady value of approximately $10\text{s}$ after the fault. It should be noted that the total aggregate rating of the two SynCos in this case is smaller than that of the single SynCo used in the previous experiment to stabilize the system. 

\begin{figure}[h]   
	\centering   
	\includegraphics[width=.75\linewidth]{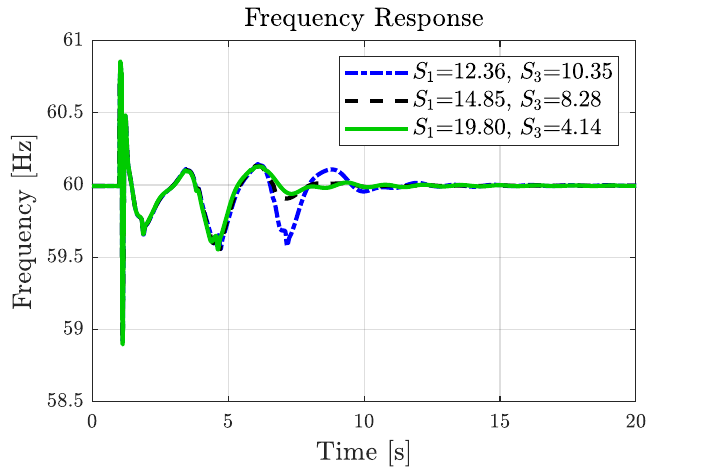}   
	\caption{Frequency response for cases with two SynCo - varying distribution of condensers rating - All inertial constants are $\text{H}=4{s}$}
	\label{fig:double_varying_rating} 
\end{figure}

To assess the impacts of the spatial distribution of SynCos, we varied their ratings. To this end, we incrementally increased the rating of condenser 1 and reduced the rating of condenser 3 while the inertia constants were kept unchanged at $\text{H}=4{s}$, The results are shown in Fig. \ref{fig:double_varying_rating}. 

The results show an improved dynamic performance with increased heterogeneity of condenser ratings that manifests itself in the form of faster decay of oscillations. In the initial case with two condensers having the closest matching ratings (highlighted by the blue trace in Fig. \ref{fig:double_varying_rating}), frequency reached a settling value after 10 seconds. The best response came from the case with the largest rating mismatch (distinguished by green trace in in Fig. \ref{fig:double_varying_rating}) in which the frequency settled in only $7.5\text{s}$ with minimal swings. Our observation here suggests the installation of SynCos at scale will require careful study to determine their optimal rating and placement. 

\begin{figure}[h]   
	\centering   
	\includegraphics[width=.75\linewidth]{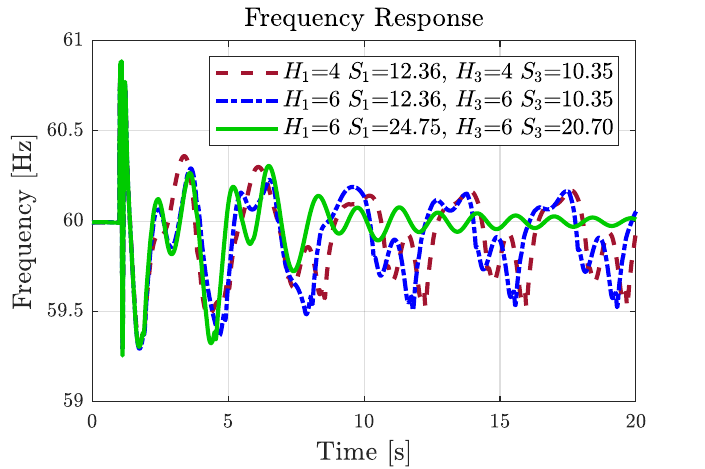}     \caption{Frequency response for low-inertia cases with two SynCo - different condenser ratings with homogeneous inertia constants.}
	\label{fig:double_low_inertia_homogeneous_varying_capacity} 
\end{figure}

Let us focus on our initial stable case with two SynCos, shown as the green trace in Fig. \ref{fig:double_base}. Reducing inertia constant of the only synchronous generator of this case from $\text{H}=4{s}$ to $\text{H}=2{s}$ resulted in system's loss of stability. Subsequently, we attempted to restore the stability by increasing the inertia constants of both SynCos from $\text{H}=4{s}$ to $\text{H}=6{s}$ without changing their rating, keeping them at $\text{S}_1=12.36\text{ MVA}$ and $\text{S}_3=10.35 \text{ MVA}$, but it was insufficient. Increasing the condensers respective ratings to $\text{S}_1=24.75\text{ MVA}$ and $\text{S}_3=20.70 \text{ MVA}$, eventually stabilized this system. The results are shown in \ref{fig:double_low_inertia_homogeneous_varying_capacity}.

\begin{figure}[h]   
	\centering   
	\includegraphics[width=.75\linewidth]{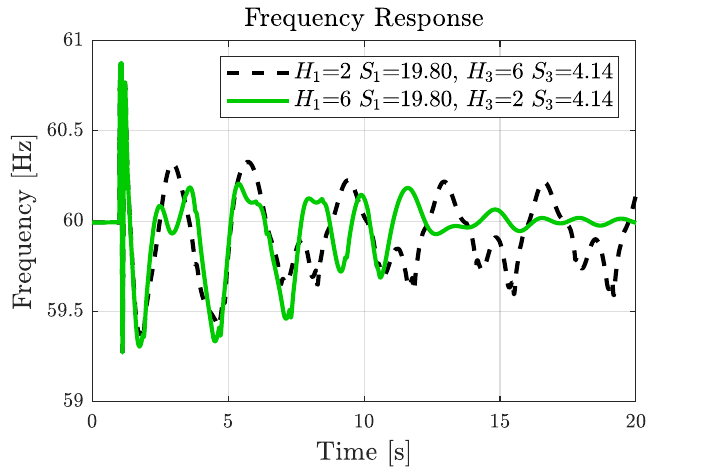}     \caption{Frequency response for low-inertia cases with two SynCo - different condenser ratings with heterogeneous inertia constants.}
	\label{fig:double_low_inertia_homogeneous} 
\end{figure}

Having learnt that the heterogeneity of SynCos ratings may enhance the dynamic response, next we tried to stabilize the latest case in which the synchronous generator's inertia constant was reduced to $\text{H}=2{s}$ by taking advantage of inertia and rating distribution, examining two cases. The first case had a large condenser with a small inertia constant along with a small condenser with a large inertia constant. The second case had a large condenser with a large inertia constant along with a small condenser with a small  inertia constant. The results are shown in Fig. \ref{fig:double_low_inertia_homogeneous}.

These results show that the second case became stable (shown in the green trace in Fig. 
\ref{fig:double_low_inertia_homogeneous}), that was installing a large SynCo with high inertia constant along with a small SynCo with low inertia constant. Interesting to note that the total aggregate rating of the two SynCos in this case (that is $23.94\text{ MVA}$) is still smaller than that of the single device we used in our experiment to stabilize the system relying solely on one SynCo (that was $24.75\text{ MVA}$), shown in Fig. \ref{fig:freqs_C1}. 

Overall, our results suggest that the placement of SynCos requires a careful consideration of the system's phase angle, frequency, and/or voltage challenges, needs, and characteristics in order to appropriately identify their capacity, inertia constant, and location. If SynCos are poorly planned or allocated, they could have adverse effect on system dynamics and even destabilize the system.

\section{Conclusion}

SynCos have historically been used in power systems as essentially "\textit{power factor correction}" assets. With the proliferation of grid following inverter-based resources that, when installed at large-scale, may create "\textit{low-inertia}" and "\textit{weak}" operating conditions, SynCos are experiencing a renaissance. This has prompted a reanalysis of SynCos and their potential to play new roles. To summarize the findings:

\begin{itemize}
	
	\item SynCos are assets that help enhance the "\textit{strength}" of the grid but our interpretation yields that their contribution does not meet the requirement for \textit{forming}" the grid. They have a remarkable capabilities to help the grid with dynamic response, but because they intrinsically synchronize with the frequency of the connected grid, they fall short of our definition of grid forming capabilities in terms of their lack of independence in grid phase angle and frequency construction. Their synchronization with the interconnected grid is a mechanical analog of PLLs in inverters. 
	
	\item SynCos provide frequency response support because of the kinetic energy stored in their rotor, directly proportional to the momentum of the SynCo. When disturbed, the larger the momentum (or inertia), the longer the rotor will continue to spin as it slows down. Therefore, the longer the phase angle continues to evolve - and grid formation continues - the slower the frequency decline. When paired with a GFL, it will support the GFL operation by slowing frequency deviations, making them easier for the GFL's PLL to track. This capability is particularly prevalent in operating conditions when inertia is low.
	
	\item Because a SynCo is a doubly excited machine, during a fault it transitions into a synchronous generator. Subsequently, for a short period of time, it can supply short-circuit current, that is composed of active and reactive current. This feature becomes very practical in the system where rotor angle stability is an issue. When paired with GFLs, which have low fault current contributions, this is a very important feature because it helps with fault detection. Since the low fault current contribution of GFLs is the product of the semiconductor switches in inverters, it is also a limitation of grid-forming control. The supply of fault current may thus be the key contribution of SynCos in future power grids. It also helps with settling post-fault oscillations by providing damping. 
	
	\item The outstanding capability of SynCo to regulate voltage in steady-sate, prevent large voltage swings, and damp out voltage oscillations, helps improve the system response following a disturbance. This is a crucial feature in low-inertia/weak grids that aid operation and synchronization of GFLs. 
	
	\item GFLs rely on a PLL to operate. Installation of SynCos can be viewed as a means to improve the availability of a good quality voltage waveform in both intact and transient operation that then the PLL of GFLs can use to continue injecting power. The voltage magnitude is improved by electromagnetic components of the SynCo whereas the voltage phase and, subsequently, frequency, are improved by its electromechanical components.

\end{itemize}

Beyond the scope of this research, there remain major questions about network topology and spatial and capacity distribution of SynCos and GFLs to better understand the interactive dynamics they may manifest in large-scale systems. The fact is that no amount of SynCo will be enough to enable and guarantee a stable system if system planning does not provide sufficient generation resources and optimally located grid-forming assets.







\end{document}